\documentclass[aps,prl,twocolumn,showpacs]{revtex4}

\usepackage{graphicx}

\begin{document}

\title{Core Size Effect on the Vortex Quasiparticle Excitations in Overdoped $La_{2-x}Sr_xCuO_4$ Single Crystals}

\author{Z. Y. Liu$^1$, H. H. Wen$^1$\email, T. Xiang$^{2,1}$, Seiki Komiya$^3$, X. F. Sun$^3$, Yoichi Ando$^3$}

\affiliation{$^1$ National Laboratory for Superconductivity,
Institute of Physics, Chinese Academy of Sciences, P.~O.~Box 603,
Beijing 100080, P.~R.~China}

\affiliation{$^2$ Institute of Theoretical Physics and ICTS,
Chinese Academy of Sciences, P. O. Box 2735, Beijing 100080,
P.~R.~China}

\affiliation{$^3$ Central Research Institute of Electric Power
Industry, Komae, Tokyo 201-8511, Japan}

\date{\today}

\begin{abstract}
Low temperature specific heat has been measured for an overdoped
$La_{2-x}Sr_xCuO_4$ ( x = 0.22 ) single crystal. The quasiparticle
density of states ( DOS ) in the mixed state is found to deviate
from the predicted scaling law $C_{vol}=Hf(T/H^{1/2})$. However,
this scaling behavior is nicely reconciled if one considers the
normal core region ( $\xi \approx$ 21$\AA$ ) which gives small
contribution to the total DOS. The radius of the core and other
parameters derived are consistent with reported values. Our
results suggest that there is no zero-bias conductance peak ( ZBCP
), which is predicted by the simple Bogoliubov de-Gennes theory in
the vortex core of a d-wave superconductor.
\end{abstract}

\pacs{74.20.Rp, 72.25.-q, 74.25.Fy, 74.72.Dn}

\maketitle One of few points with consensus in the cuprate
superconductors is the $d_{x^2-y^2}$ pairing symmetry in hole
doped cuprates. This has been supported by tremendous experiments
\cite{Tsuei1} both from surface
detection\cite{Tsuei2,ARPES,Hardy,Tunneling,YehNC} and bulk
measurements\cite{NMR,Moler,Revaz,Wright}. In a d-wave
superconductor with line nodes in the gap function, the
quasiparticle density of states ( DOS ) $N(E)$ rises linearly with
energy at the Fermi level in zero field, $N(E)\propto|E-E_F|$,
resulting in an electronic specific heat $C_e=\alpha T^2$
\cite{Kopnin1996}, where $\alpha = 2.52\gamma_n/T_c$ and
$\gamma_n$ is the specific heat coefficient reflecting the DOS at
the Fermi level of normal state. In the mixed state with the field
higher than a certain value, the DOS near the Fermi surface
becomes finite, therefore the quadratic term $C_e=\alpha T^2$ will
be surpassed and substituted by both the localized excitations
inside the vortex core and the de-localized excitations outside
the core. Volovik \cite{Volovik} pointed out that for d-wave
superconductors in the mixed state, supercurrents around a vortex
core lead to a Doppler shift to the quasi-particle excitation
spectrum, which affects strongly the low energy excitation around
the nodes. It was shown that the contribution from the delocalized
part will prevail over the core part and the specific heat is
predicted to behave as $C_{vol}=k\gamma_nT\sqrt{H/H_{c2}}$ with
$k$ a parameter in the order of unity. This prediction has been
verified by many specific heat measurements which were taken as
the evidence for d-wave pairing
symmetry\cite{Moler,Revaz,Wright,Phillips}. In the finite
temperature and field region a scaling law is
proposed\cite{SimonLee}

\begin{equation}
\frac{C_{vol}}{H}=f(\frac{T}{\sqrt{H}})
\end{equation}
where $f$ is a unknown scaling function. This scaling law has been
proved in $YBCO$\cite{Revaz,Wright} and in optimally doped
LSCO\cite{Phillips}. It remains however unclear whether this
scaling law is still valid in the overdoped region.

Another important but controversial issue is the vortex core state
in the cuprate superconductors. By solving the mean-field
Bogoliubov-de Gennes ( BdG ) equation, theoretically it is
suggested that a zero-bias conductance peak ( ZBCP ) exists in the
vortex core\cite{WangYong,Franz}. However this is in sharp
contrast with the experimental
observations\cite{Maggio,Renner,Pan,Hoogenboom,Dagan,Mitrovic}.
The absence of ZBCP was attributed to the presence of $id_{xy}$ or
$is$ components\cite{Dagan}. In this Letter we show that the DOS
due to vortex quasi-particle excitations deviates from Simon-Lee
scaling law, but it can be nicely reconciled if the vortex core
size is taken into account. Furthermore a consequence from our
analysis is the absence of ZBCP in the vortex core and the
disappearance is intrinsic rather than due to the second component
of order parameters.

\begin{figure}
\includegraphics[width=8cm]{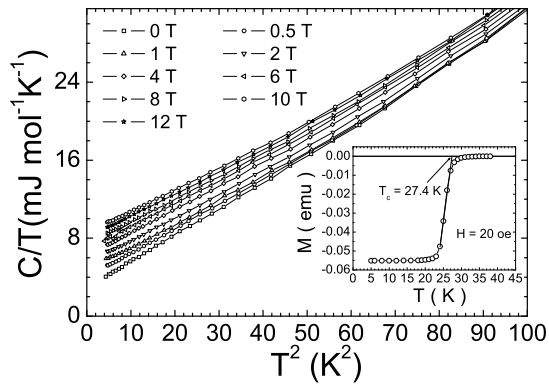}
\caption{Specific heat coefficient C/T vs. $T^2$ at magnetic
fields ranging from 0 to 12 T for the overdoped single crystal.
The inset shows the diamagnetic transition at around 27.4 K
determined by the crossing point of the extrapolating line of the
most steep part with the normal state background M = 0. }
\label{fig1}
\end{figure}

The single crystals measured in this work were prepared by the
travelling solvent floating-zone technique. In this Letter we
present data of an overdoped single crystal ( x = 0.22, $T_c =
27.4 K$) as characterized by AC susceptibility and DC
magnetization ( shown by the inset in Fig.1 ). The quality of our
sample has also been characterized by x-ray diffraction patterns
showing only (00l) peaks, and resistive measurement showing a
rather narrow transition temperature $\Delta T_c \leq $ 2 K. A
piece about 28.55 mg in mass, $3.66 \times 2.3 \times 0.5 mm^3$ in
dimension, was chosen for the specific heat measurement. The heat
capacity data presented here were taken with the relaxation method
\cite{Bachmann1972} based on an Oxford cryogenic system Maglab.
The heat capacity is determined by a direct measurement of the
thermal time constant, $\tau=(C+C_{add})/\kappa_w$, here $C$ and
$C_{add}$ are the heat capacity of the sample and addenda (
including a small sapphire substrate, small printed film heater,
tiny Cernox temperature senser, $\phi$25 $\mu m$ gold wire leads,
Wakefield thermal conducting grease ( $100\mu g$ ) ) respectively,
where $\kappa_w$ is the thermal conductance between the chip and a
thermal link. The value $C_{add}$ has been measured and subtracted
from the total capacitance, thus $C$ value reported here is only
the capacitance of the sample. We have also checked the field
dependence of $C_{add}$ and found that the change ( if any ) of
$C_{add}$ under 12 T is in the same order of the noise background
here ( 20 nJ/K at 5 K and 40 nJ/K at 20 K ). The influence of the
magnetic field ( 12 T ) on the readout of the thermometer is below
0.02 K and can be neglected. During the measurement the field was
applied parallel to c-axis and the sample was cooled to the lowest
temperature under a magnetic field (field-cooling) followed by
data acquisition in the warming up process.

Fig.1 shows the specific heat coefficient $C/T$ as a function of
$T^2$ in magnetic fields ranging from 0 to 12 T. The separation
between each field can be well determined. In low temperature
region the curves are rather linear showing that the major part is
due to phonon contribution $C_{ph}=\beta T^3$. In addition the
curve at zero field extrapolates to a finite value ( $\gamma_0$ )
at 0 K instead of zero as observed in other cuprate
superconductors. This may be interpreted as potential scattering
to the vanishing gaps near the node of $d_{x^2-y^2}$ gap function
due to small amount impurities\cite{Moler}. As also observed by
other groups for $La-214$ system, the anomalous upturn of $C/T$
due to the Schottky anomaly of free spins is very
weak\cite{Phillips}. This avoids the complexity in the data
analysis. It is known that the phonon part has a very weak field
dependence, this allows to remove the phonon contribution by
subtracting the $C/T$ at a certain field with that at zero field.
The results after the substraction are shown in Fig.2. The
subtracted values $\Delta \gamma=\gamma_H-\gamma_0 = C(T,
H)/T-C(T, H=0)/T$ show a rather linear $T$ behavior in low
temperature region ( below 8 K ) as indicated by the solid lines.
Our analysis is based on the data below 12 K. In the following we
will show that the field dependent slope of the linear part in low
temperature region shown in Fig.2 directly deviates from the
Simon-Lee\cite{SimonLee} scaling law ( eq.1 ).

\begin{figure}
\includegraphics[width=8cm]{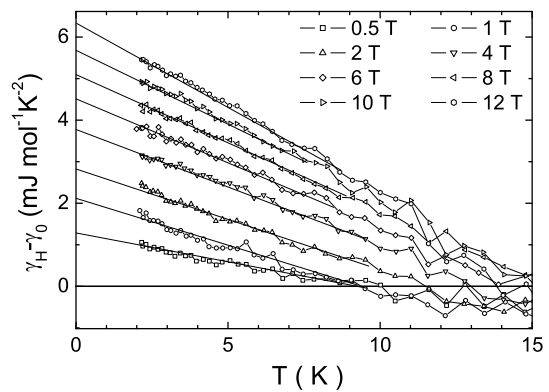}
\caption{Temperature dependence of the subtracted specific
coefficient $\gamma_H-\gamma_0 = C/T(H)-C/T(H=0)$. A linear
behavior is clearly seen in low temperature region which is not in
accord with the proposed scaling law (eq.1) by Simon and Lee ( see
text ). The linear lines in low temperature region are guides to
the eyes. From these lines one can determine the zero temperature
intercept $\Delta\gamma$ and the slope $d\gamma/dT$ shown in Fig.3
} \label{fig2}
\end{figure}

According to the scaling law, the low temperature expansion of
right hand side reads

\begin{equation}
C_{vol}=b_0H+b_1T\sqrt{H}+b_2T^2+o(T^3)
\end{equation}

where $b_0$=0 because $C_{vol}/T$ should be finite when $T=0$ and
$H\neq 0$. Since $o(T^3)$ is very small in low temperature region,
one has $C_{vol}/T=b_1\sqrt{H} + b_2 T$. When $H_{c1} << H <<
H_{c2}$, the total specific heat contains four parts: Doppler
shift term from the region outside the core $C_{vol}$, the inner
vortex core term $C_{core}\propto HT$, the small impurity
scattering term $\gamma_0T$ and the phonon term $C_{ph}$. At zero
field, the total specific heat contains three parts: the small
impurity scattering term $\gamma_0T$ and the phonon term $C_{ph}$
both depend on magnetic field weakly, and a quadratic term $\alpha
T^2$ due to the thermal excitation near the nodal region. Thus
$\Delta \gamma$ can be written as:

\begin{equation}
\Delta \gamma=\gamma_H-\gamma_0=b_1\sqrt{H}+(b_2-\alpha
)T+\gamma_{core}H
\end{equation}

From eq.3 it is clear that $\Delta \gamma$ depends on T through
the second term, however the slope $b_2-\alpha$ is $H$ independent
according to the definition. This clearly indicates that the
Simon-Lee scaling law cannot be directly applied to interpret the
field dependent slope of $\Delta \gamma$ vs. $T$ as shown in
Fig.2.

In order to understand the underlying physics, still based on the
Simon-Lee scaling law, we propose that the core size effect has a
sizable influence on the total vortex quasi-particle excitations.
By taking account the vortex core size ($2\xi$), one can rewrite
$\Delta \gamma$ as:

\begin{equation}
\Delta \gamma=(b_1\sqrt{H}+b_2T)\times(1-\xi^2/R_a^2)-\alpha
T+\gamma_{core}H
\end{equation}

where $\xi$ is the radius of the normal core, $R_a$ is the radius
of a single vortex $R_a^2=\phi_0/\pi H$. Thus eq.4 can be written
as:

\begin{equation}
\Delta \gamma=b_1\sqrt{H}\times(1-\frac{\pi
\xi^2}{\phi_0}H)+(b_2-\alpha
)T-b_2\frac{\pi\xi^2}{\phi_0}HT+\gamma_{core}H
\end{equation}

One immediately realizes that the third term in eq.5 is just what
we need for interpreting the difficulty as mentioned above. Next
let us have a closer inspection at the data and derive some
parameters. At zero temperature, only the first term and the last
term are left. The values of $\Delta \gamma(T=0)$ are determined
from the extrapolation of the linear lines in Fig.2 to 0 K and
presented in Fig.3. The solid line is a fit to the data using the
first term in eq.5 yielding  $b1=1.9 \pm 0.042 mJK^{-2}T^{-1/2}$
and $\pi\xi^2/\phi_0=0.0067\pm0.002$ and thus $\xi=21\AA$. The
value $\xi=21\AA$ derived here is quite close to that found in
Nernst\cite{WangYaYu} and STM measurements\cite{Pan}. We also
tried to use the first term together with the last term to fit the
data but find out that the contribution from the last term is
extremely small. The first term here describes the zero
temperature data very well, indicating the absence of a bulk
second order parameter such as $id_{xy}$ or $is$ since otherwise
the nodal point would be filled completely and the Doppler shift
had very weak effect on the quasi-particle excitations. The inset
of Fig.3 shows the field dependence of the slope of the linear
part in Fig.2. It is clear that the slope increases with H above 1
T. This can be exactly anticipated by the second and third terms
in eq.5. From the inset of Fig.3 one obtains $\alpha-b_2=0.21mJ
mol^{-1} K^{-3}$ and $b_2\pi\xi^2/\phi_0=0.015 mJ
mol^{-1}K^{-3}T^{-1}$. By taking $\xi = 21 \AA$, we obtain the
following values: $\alpha=2.449 mJ mol^{-1}K^{-3}$ and
$b_2=2.239mJ mol^{-1}K^{-3}$. It is known that
$\alpha=2.52\gamma_n/T_c$, we further obtain $\gamma_n = 26.63 mJ
mol^{-1}K^{-2}$. In addition, it is predicted that
$b_1=k\gamma_n/\sqrt{H_{c2}}$ which yields also $\gamma_n = 23.3
mJ mol^{-1}K^{-2}$ if taking $k=1$ and $H_{c2}=\phi_0/\pi\xi^2$.
Two different approaches lead to rather close values of
$\gamma_n$. This value is also close to that found in
$Y-123$\cite{Moler}.

The nice fitting in Fig.3 with only the first term of eq.5
suggests that the core region has very small contribution to the
DOS since otherwise the last term $\gamma_{core}H$ should be
sizeable. This implies that the low energy DOS inside the vortex
core is very small, ruling out the presence of ZBCP as expected by
the BdG theory for a d-wave superconductor\cite{WangYong,Franz}.
This conclusion is further strengthened in the following by the
fact that the core size effect has a sizeable influence on the DOS
due to Doppler shift, but the contribution from inner part of the
core is small. In order to further test our idea ( eq.5 ), we
present the scaling of $\Delta\gamma T -(b_2-\alpha)T^2$ vs.
$T\sqrt{H}$ in Fig.4 with $b_2-\alpha$ derived above. The quality
of our scaling is remarkable. This can be easily understood from
eq.5 when the correction $\pi\xi^2H/\phi_0 ( = 0.08 << 1$ at 12 T
) to the first term is very small. The slight scattering in the
high temperature and field region is due to the noise of the data.
Worthy of noting is that to have this nice scaling we need to take
$\gamma_{core}\approx 0$, again showing a small contribution from
the inner vortex core. The solid line is a theoretical curve from
eq.5 by using the parameters derived above. Both the nice scaling
and the consistency between the experimental data and eq.5 suggest
that the Simon-Lee scaling law can be reconciled by considering
the vortex core size effect. Meanwhile the scaling according to
the original Simon-Lee theory is also presented in the inset of
Fig.4. showing a rather poor scaling quality.

\begin{figure}
\includegraphics[width=8cm]{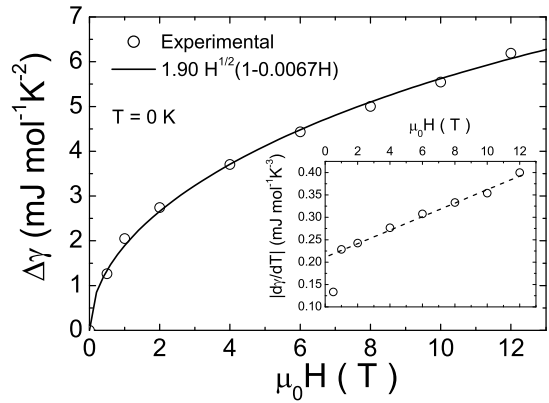}
\caption{DOS induced by Doppler shift at zero K. The solid line is
a theoretical curve $\Delta\gamma=1.9\sqrt{H}(1-0.0067H)$. The
remarkable consistency between data and theory based on
$d_{x^2-y^2}$ symmetry indicates that we don't have any second
component such as $id_{xy}$ or $i$s at fields ranging from 0 to 12
T. The inset shows the slope $d\gamma/dT$ of the straight lines
shown in Fig.2 in low temperature region. The dashed line is a
linear fit to the data at fields above one tesla. The intercept
and the slope of the dashed line give rise to the pre-factors of
the second and third terms in eq.5} \label{fig3}
\end{figure}

\begin{figure}
\includegraphics[width=8cm]{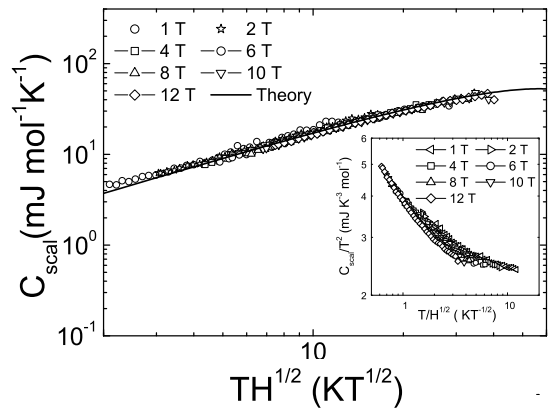}
\caption{ Scaling of the data
$C_{scal}=C(H)-C(H=0)+(\alpha-b_2)T^2$ vs. $T\sqrt{H}$. The solid
line is a theoretical fit according to eq.5 with fitting
parameters derived in the text. A perfect scaling and a nice
consistency between data and the scaling law with core-size
correction is clear. The inset shows the scaling according to
Simon and Lee ( eq.1 ) where $C_{scal}=C(H)-C(H=0)+\alpha T^2$. }
\label{fig4}
\end{figure}

In low temperature region, our analysis indicates that the field
induced DOS can be well described by Volovik's theory or Simon-Lee
scaling law although a correction due to the core size effect is
needed. This means that the prerequisite for the theory, i.e., the
$d_{x^2-y^2}$ pairing symmetry is well satisfied. Therefore it
naturally rules out the possible presence of a second order
parameter like $id_{xy}$ or $is$ either due to
overdoping\cite{YehNC} or due to the field effect\cite{Laughlin}
in the present overdoped sample. Meanwhile another seemingly
contradicted phenomenon is that the vortex core region contributes
very little ( at least much smaller than that induced by the
Doppler shift if the super-current would flow in the same area )
to the total DOS. This suggests that there is no ZBCP within the
normal core since otherwise the normal core region should give
rise to a sizable signal. Our conclusion is consistent with the
tunnelling results\cite{Maggio,Renner,Pan,Hoogenboom,Dagan} and
certainly clears up the concerns about the surface conditions in
the STM measurement since our specific heat data reflect the bulk
property. Recent results from NMR also show the absence of ZBCP
inside the vortex core\cite{Mitrovic}. In this sense our data
together with the earlier NMR data present a bulk evidence for an
anomalous vortex core. Interestingly it is widely perceived that
the normal state in overdoped region shows a Fermi liquid behavior
even when the superconductivity is completely
suppressed\cite{Proust}. Clearly the mean-field frame of BdG
theory based on the conventional d-wave superconductivity is not
enough to interpret the anomalous vortex core state.

In conclusion, the quasiparticle density of states ( DOS ) due to
Doppler shift in the mixed state of an overdoped
$La_{2-x}Sr_xCuO_4$ ( x = 0.22 ) single crystal is found to
deviate from the proposed scaling law of Simon and Lee. However,
this law is reconciled if one considers the core size effect . The
contribution from the inner vortex core is small comparing to that
due to the Doppler shift in the same area. Our results suggest the
absence of the ZBCP in the vortex core although it is expected by
the Bogoliubov de-Gennes theory for a d-wave superconductor.

\section{Acknowledgments}

This work is supported by the National Science Foundation of China
(NSFC 19825111, 10174090, 10274097), the Ministry of Science and
Technology of China ( project: NKBRSF-G19990646 ), the Knowledge
Innovation Project of Chinese Academy of Sciences. We are grateful
for fruitful discussions with D. H. Lee, P. C. Dai, S. H. Pan and
Y.Y. Wang.


Correspondence should be addressed to hhwen@aphy.iphy.ac.cn

\end{document}